\documentclass[conference]{IEEEtran}
   \usepackage[pdftex]{graphicx}
\usepackage{amsmath}
\DeclareMathSizes{10}{10}{8}{7}   

\begin{document}
%
\title{Capacity Scaling Laws for Underwater Networks }


\author{\authorblockN{Daniel E. Lucani}
\authorblockA{RLE, MIT\\
Cambridge, Massachusetts, 02139\\
Email: dlucani@mit.edu}
\and
\authorblockN{Muriel M\'edard}
\authorblockA{RLE, MIT\\
Cambridge, Massachusetts, 02139\\
Email: medard@mit.edu}
\and
\authorblockN{Milica Stojanovic}
\authorblockA{Northeastern University\\
Boston, Massachusetts, 02115\\
Email: millitsa@mit.edu}}


%

\specialpapernotice{(Invited Paper)}

\maketitle

\begin{abstract}
The underwater acoustic channel is characterized by a path loss that depends
not only on the transmission distance, but also on the signal
frequency. Signals transmitted from one user to another over a distance $l$ 
are subject to a power loss of $l^{-\alpha}{a(f)}^{-l}$. Although a terrestrial radio channel can be modeled similarly, the underwater acoustic channel has different characteristics. The spreading factor $\alpha$, related to the geometry of propagation, has values in the range $1 \leq \alpha \leq 2$. The absorption coefficient $a(f)$ is a rapidly increasing function of frequency: it is three orders of magnitude greater at 100~kHz than at a few Hz. Existing results for capacity of wireless networks correspond to scenarios for which $a(f) = 1$, or a constant greater than one, and $\alpha \geq 2$. These results cannot be applied to underwater acoustic networks in which the attenuation varies over the system bandwidth. We use a water-filling argument to assess the minimum transmission power and optimum transmission band as functions of the link distance and desired data rate, and study the capacity scaling laws under this model. 
\end{abstract}


%
\IEEEpeerreviewmaketitle

\section{Introduction}

	The seminal work by \cite{GuptaKumar00} studied wireless networks, modeled as a set of $n$ nodes that exchange information, with the aim of determining what amount of information the source nodes can send to the destination as the number $n$ grows. The original results obtained for nodes deployed in a disk of unit area motivated the study of capacity scaling laws in different scenarios, ranging from achievability results in random deployments using percolation theory \cite{Franceschetti07} or cooperation between nodes \cite{Ozgur07}, to the impact of node mobility over the capacity of the network, e.g. \cite{Grossglauser02}. Reference \cite{Vu08} provides a good overview of the different assumptions and scaling laws for radio wireless networks.  
	
	The underwater acoustic channel is characterized by a path loss that depends not only on the transmission distance, but also on the signal frequency \cite{milica06}. Signals transmitted over a distance $l$  
are subject to a power loss of $l^{-\alpha}{a(f)}^{-l}$. Although a terrestrial radio channel can be modeled similarly, the underwater acoustic channel has different characteristics. The spreading factor $\alpha$, related to the geometry of propagation, has values in the range $1 \leq \alpha \leq 2$, where $\alpha = 1$ corresponds to cylindrical spreading. Also, the absorption coefficient $a(f)$ is a rapidly increasing function of frequency, e.g. it is three orders of magnitude greater at 100~kHz than at a few Hz \cite{milica06}. Finally, the power spectral density of the noise underwater is highly dependent on frequency.    
	
	Existing capacity scaling laws for wireless radio networks correspond to scenarios for which $a(f) = 1$, or a constant greater than one, and $\alpha \geq 2$, e.g. \cite{GuptaKumar00}, \cite{Franceschetti07}. These results cannot be directly applied to underwater acoustic networks in which the attenuation varies over the system bandwidth and $\alpha\leq 2$.
We study the scaling laws under a model that considers a water-filling argument to assess the minimum transmission power and optimum transmission band as functions of the link distance and desired data rate \cite{lucani08model}. In particular, we study the case of arbitrarily deployed networks in a disk of unit area, and follow a similar procedure as in \cite{GuptaKumar00} to derive an upper bound on capacity. In this sense, we provide an extension of the work in \cite{GuptaKumar00} under a more complicated power loss model. 

We show that the amount of information that can be exchanged by each source-destination pair in underwater acoustic networks  goes to zero as the number of nodes $n$ goes to infinity. This occurs at least at a rate $n^{-1/\alpha}e^{-W_0(O(n^{-1/\alpha}))}$, where $W_0$ represents the branch zero of the Lambert function. 
We illustrate that this throughput per source-destination pair has two different regions. For small $n$, the throughput decreases very slowly as $n$ increases. For large $n$, it decreases almost as $n^{-1/\alpha}$. Thus for large enough $n$, the throughput decreases more rapidly in underwater networks than in typical radio channels, because of the difference in the path loss exponent $\alpha$.    
		
	 The paper is organized as follows. In Section II, we present the underwater channel model. In Section III, we analyze the scaling laws for the case of a network transmitting in an arbitrarily chosen narrow band. In Section IV, we study scaling laws for the low-power/narrow-band case, with optimal bandwidth allocation using a waterfilling argument. In section V, we consider the case in which the nodes can transmit at high power over a wide transmission band. Conclusions are summarized in Section VI. 


\section{Underwater Channel Model}
An underwater acoustic channel is characterized by an attenuation that depends on the distance \textit{l} and the signal frequency \textit{f} as
\begin{eqnarray} \label{PathLoss}
A(l,f) = \left( \frac{l}{l_{ref}} \right)^{\alpha} {a(f)}^{l}
\end{eqnarray} 
where $l_{ref}$ is a reference distance (typically 1 m). 

A common empirical model used for the absorption $a(f)$ is Thorp's formula \cite{milica06} which captures the dependence on the frequency. This absorption $a(f)$ is an increasing function of $f$. The spreading factor describes the geometry of propagation and is typically $1\leq\alpha\leq2$, e.g. $\alpha=1$ and $\alpha =2$ correspond to cylindrical and spherical spreading, respectively.  
	The noise in an acoustic channel can be modeled through four basic sources: turbulence, thermal noise, shipping, and waves. It has a power spectral density (psd) which depends on the frequency, the shipping activity $s$, and the wind speed $w$ in m/s \cite{milica06}.

The complete model for a colored Gaussian underwater link was presented in \cite{lucani08model} where power was allocated through waterfilling. In the absence of multipath and channel fading, the relationship among capacity, transmission power, and optimal transmission band of a point-to-point link is given by \cite{lucani08model}
\begin{align} \label {Capacity_func}
C = \int _{B(l,C)} \log_{2} \left ( \scriptstyle  \frac{K(l,C)}{A(l,f)N(f)} \displaystyle  \right ) df
\end{align} 
where $N(f)$ is the psd of the noise, $B(l,C)$ is the optimum band of operation and $K(l,C)$ is a constant. 
The transmission power associated with a particular choice of $(l,C)$ is given by
\begin{align} \label{Power_func}
P(l,C) = \int _{B(l,C)} S(l,C,f) df
\end{align} 
where the psd of the signal is $ S(l,C,f) = K(l,C) - A(l,f)N(f), f \in B(l,C)$. 

\begin{figure}[t]
\centering	
\includegraphics[height=2.3in,width=3.3in, keepaspectratio]{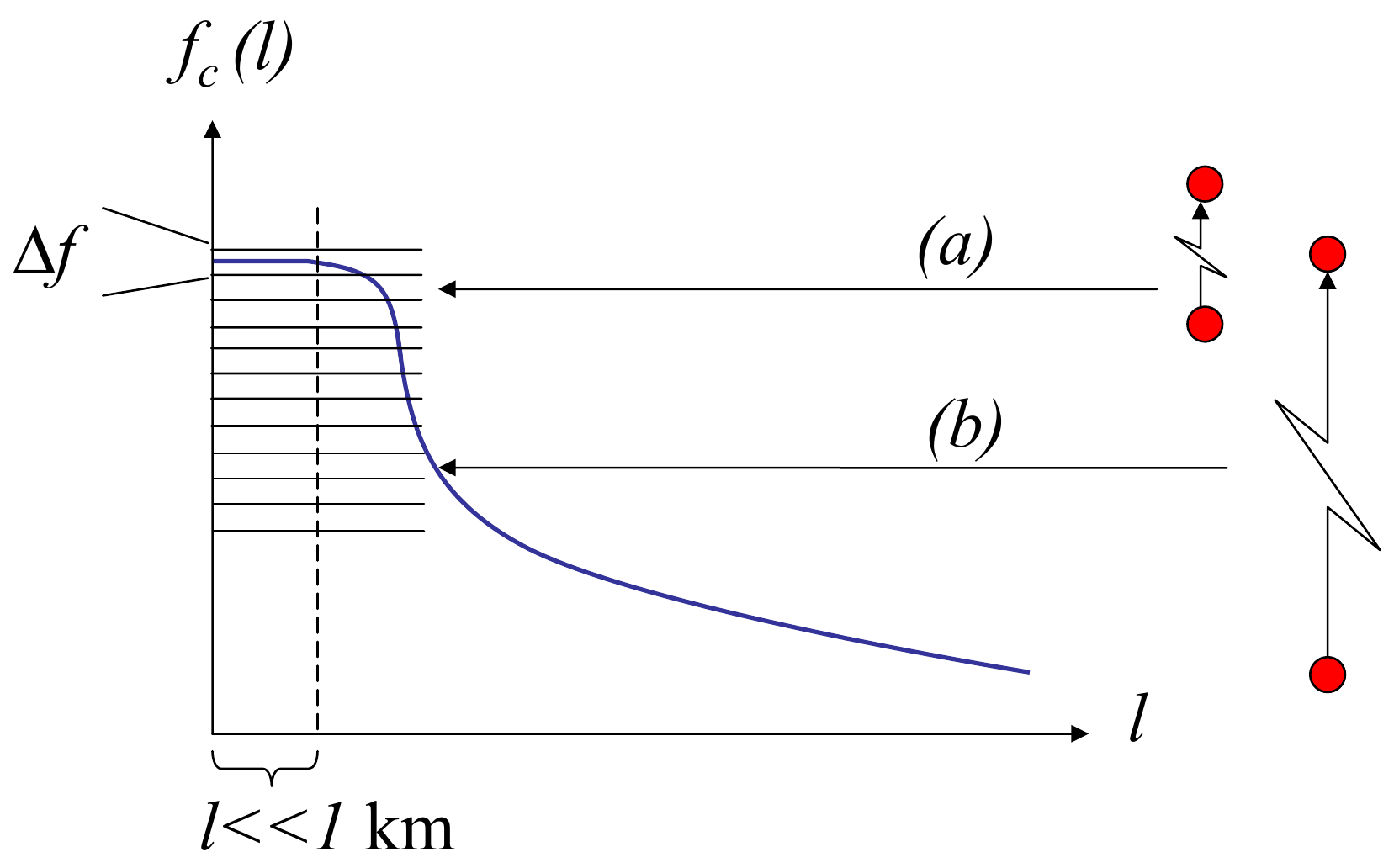}
\caption{Relationship between transmission distance and center frequency in a narrow band system.}
\label{Distance_Band.tag}
\end{figure}

A distinguishing feature of the underwater acoustic channel is the dependence of the optimal transmission band on the link distance \cite{lucani08model}. Fig. 1 illustrates the optimal center frequency $f_c(l)$ as a function of distance. The optimal center frequency is defined as the frequency at which $A(l,f)N(f)$ is minimal. This implies that if the transmission power for a link is low, the transmission bandwidth will be low and around the optimal frequency. Thus, the optimal transmission band in the spectrum changes dramatically with the link distance. Fig.~\ref{Distance_Band.tag} also illustrates that a node transmitting over a short range will optimally be assigned a transmission band at high center frequency, as in case (a), while a node transmitting over a longer distance will be assigned a different transmission band at lower center frequency, as in case (b).

For the case in which power available for transmission is low, the bandwidth of the transmission band will also be small. 
When the bandwidth is low enough $|B(l,C)| = \Delta f$, such that the product $A(l,f)N(f)$ does not change much over that band, one can make a Taylor series approximation around the center frequency $f_c(l)$. This allows us to determine the power $P$ for which the transmission band is narrow due to our waterfilling argument. The Taylor series approximation has the form
\begin{eqnarray}\label{AN_approx}A(l,f)N(f) \approx A(l,f_c)N(f_c) + \Upsilon  \frac{{(f - f_c)}^2}{2}\end{eqnarray}
$\forall f$ in the band, where $\Upsilon = {\frac{\partial^2}{\partial f^2} \left( A(l,f)N(f)\right)}\mid_{f = f_c}$.   
Substituting the expression~\eqref{AN_approx} into expression~\eqref{Power_func}, and using the fact that $K(l,C) = A(l,f_{\max})N(f_{\max}) = A(l,f_{\min})N(f_{\min})$, where $f_{\max}$ and $f_{\min}$ are the maximum and minimum frequencies of the transmission band, we obtain
\begin{eqnarray}
&P(l,C) \approx A(l,f_{\max})N(f_{\max}) \Delta f\notag \\&- \int _{f_{\min}}^{f_{\max}} \left( A(l,f_c)N(f_c) + \Upsilon  \frac{{(f - f_c)}^2}{2} \right) df
\end{eqnarray}
where $\Delta f = f_{\max} - f_{\min}$. Considering $f_{\max} - f_c \approx \frac{\Delta f}{2}$ and $f_c - f_{\min} \approx \frac{\Delta f}{2}$ given our quadratic Taylor series approximation of $A(l,f)N(f)$, the above expression reduces to
\begin{eqnarray}
P = \frac{\Upsilon}{12} {\Delta f}^3
\end{eqnarray}

\section{Fixed Narrow-band model}
Let us study the physical model \cite{GuptaKumar00} to obtain an upper bound to transport capacity for transmissions in a arbitrarily chosen narrow band in an underwater channel. The narrow band assumption allows us to consider the attenuation as a constant over that band. We assume that the nodes are arbitrarily deployed in a disk of unit area as in \cite{GuptaKumar00}, that each node has an intended destination node, and that the requirement for successful reception at node $j$ of a transmission from node $i$ is
\begin{eqnarray}
\frac{ \frac{P_{i}(f)}{A(|X_{i} - X_{j(i)}|,f)}  }{ N(f) + \sum_{k \in \tau, k \neq i} \frac{P_{k}(f)}{A(|X_{k} - X_{j(i)}|,f) }} \geq \beta 
\end{eqnarray} 
where $X_{i}$ is the position of node $i$, $X_{j(i)}$ is the position of node $j$ to which $i$ is transmitting, and $\tau$ is the set of all nodes transmitting simultaneously in the same transmission sub-band and time slot. We assume that all sub-bands are in the narrow band \cite{GuptaKumar00}, so that the attenuation is only dependent on the central frequency of the narrow band. The parameter $f$ is kept to keep in mind the frequency dependence, and to allow us to use these results in the following sections where we analyze more complex settings. We consider that $\lambda$ is the throughput of each node, the network transports $\lambda nT$ bits over $T$ seconds, and that the average distance between source and destination of a bit is $\bar L$. Finally, we define the transport capacity as $\lambda n \bar L$ bits-meters per second \cite{GuptaKumar00}.

Let us define $W = \triangle f \log_{2} (1 + \beta)$ to be the transmission rate, where $\triangle f$ is the bandwidth of the narrow band chosen for transmission. 
Since $|X_{k} - X_{j(i)}| \leq \frac{2}{\sqrt{\pi}}$, and $a(f)\geq 1, \forall f$, then the path loss is
\begin{equation}A(|X_{k} - X_{j(i)}|,f) \leq \left( \frac{2}{\sqrt{\pi} l_{ref}} \right)^{\alpha} {a(f)}^{2/\sqrt{\pi}} \equiv  \gamma \; .\end{equation} 
Using a similar procedure as in \cite{GuptaKumar00}, we have that
\begin{eqnarray}
A(|X_{i} - X_{j(i)}|,f) \leq \frac{\beta + 1}{\beta} \frac{\gamma P_{i}(f)}{ \sum_{k \in \tau} P_{k}(f)} \; .
\end{eqnarray} 
Let us sum over all transmitters $i \in \tau$ and use the definition of the path loss in expression~\eqref{PathLoss}:

\begin{eqnarray} \label{BreakingPoint}
\sum_{i \in \tau} {|X_{i} - X_{j(i)}|}^{\alpha} a(f)^{|X_{i} - X_{j(i)}|} \leq \gamma_{{\alpha}} \frac{\beta + 1}{\beta} 
\end{eqnarray} 
where $\gamma_{{\alpha}} = l_{ref}^{\alpha} \gamma = \left( \frac{2}{\sqrt{\pi}} \right)^{\alpha} {a(f)}^{2/\sqrt{\pi}}$.
Summing over all sub-bands and time slots and dividing both sides by $H$, we obtain
\begin{eqnarray}\label{EQ11}
\frac{1}{H}\sum_{b = 1} ^ {\lambda n T} \sum_{h = 1} ^ {h(b)} r^A (h,b) \leq \gamma_{{\alpha}} \frac{\beta + 1}{\beta} \frac{WT}{H}
\end{eqnarray} 
where $h(b)$ represents the $h$-th hop of a bit $b$, and $H$ is defined as the number of hops performed in $T$ seconds, which can be bounded by  $H \leq \frac{WTn}{2}$  \cite{GuptaKumar00}. Finally, $r^A (h,b) = {l(h,b)}^{\alpha}{a(f)}^{l(h,b)}$, where $l(h,b)$ represents the distance between receiver and transmitter for the $h$-th hop of bit $b$.
Since the function $r^A(l) = l^{\alpha} a(f)^l$ is increasing and convex for $l \geq 0$, ${\alpha} \geq 1$ and $a(f) \geq 1$, then	
\begin{equation}
\left (\frac{\ln{a(f)}}{H}\sum_{b = 1} ^ {\lambda n T} \sum_{h = 1} ^ {h(b)} l(h,b) \right )^{\alpha}  \exp{ \left (\frac{\ln{a(f)}}{H}\sum_{b = 1} ^ {\lambda n T} \sum_{h = 1} ^ {h(b)} l(h,b) \right )} \notag
\end{equation}\begin{equation}\leq (\ln{a(f)})^{\alpha} \gamma_{{\alpha}} \frac{\beta + 1}{\beta} \frac{WT}{H} \; .\end{equation}
Let us define $\psi = (\ln{a(f)})^{\alpha} \gamma_{{\alpha}} \frac{\beta + 1}{\beta} \frac{WT}{H}$, and note that $\psi \geq 0$. The left hand side of the above inequality is a Lambert function of the form $W^{\alpha} \exp{W}$, which is an increasing function when $W \geq 0$. Then,
 
\begin{eqnarray}
\frac{\ln{a(f)}}{H}\sum_{b = 1} ^ {\lambda n T} \sum_{h = 1} ^ {h(b)} l(h,b) \leq \psi^{1/{\alpha}} \exp{ \left( -W_{0}\left( \frac{\psi ^{1/{\alpha}}}{{\alpha}}\right) \right)}
\end{eqnarray}  	
where $W_{0}(\cdot)$ is the branch zero of the Lambert function, using the nomenclature in \cite{Chapeau}. This fact implies that
\begin{eqnarray}\label{PsiEq}
\lambda n \bar {L} \leq \frac{H}{T \ln {a(f)}} \psi^{1/{\alpha}} \exp{ \left( -W_{0}\left( \frac{\psi ^{1/{\alpha}}}{{\alpha}}\right) \right)} \; .
\end{eqnarray}  		

Substituting for $\psi$ in \eqref{PsiEq}, we obtain
\begin{eqnarray} \label{AlmostThere}
\lambda n \bar {L} \leq \frac{H^{\frac{{\alpha}-1}{{\alpha}}}}{T^{\frac{{\alpha}-1}{{\alpha}}} } {\left( \gamma_{{\alpha}} \frac{\beta + 1}{\beta} W \right)}^{1/{\alpha}} \exp{ \left( -W_{0}\left( \frac{\psi ^{1/{\alpha}}}{{\alpha}}\right) \right)}. 
\end{eqnarray}  	
Since $H^{\frac{{\alpha}-1}{{\alpha}}}$ is an increasing function for ${\alpha}>1$, and constant for ${\alpha}=1$, then $H^{\frac{{\alpha}-1}{{\alpha}}} \leq {\left( \frac{W T n}{2}\right)}^{\frac{{\alpha}-1}{{\alpha}}}$. Considering that $W_{0}(\cdot)$ is an increasing function, we have that
\begin{eqnarray}
W_{0}\left( \frac{\psi ^{1/{\alpha}}}{{\alpha}}\right) \geq W_{0}\left( \frac{2 \ln{a(f)} {a(f)}^{\frac{2}{{\alpha}\sqrt{\pi}}}}{{\alpha} \sqrt{\pi}}   {\left( \frac{\beta + 1}{ \beta} \right) }^{1/{\alpha}} \frac{2^{1/{\alpha}}}{n^{1/{\alpha}}} \right)\; . \notag
\end{eqnarray}

Using these inequalities into expression~\eqref{AlmostThere}, we obtain the scaling law:
\begin{eqnarray}\label{fixednarrowband}
\lambda n \bar L \le \Phi Wn^{\frac{{\alpha - 1}}{\alpha}} \exp \left( { - W_0 \left( {\Phi \frac{{2\ln a(f)}}{\alpha}\left( {\frac{1}{n}} \right)^{1/\alpha} } \right)} \right)
\end{eqnarray}  	
where
\[
\Phi  = \frac{{2^{1/\alpha} }}{{\sqrt \pi  }}\left( {\frac{{\beta  + 1}}{\beta }} \right)^{1/\alpha} \left( a(f)^{\frac{2}{{\sqrt \pi  }}} \right)^{1/\alpha} \; .
\]

Since the zero-branch of the Lambert function satisfies $W_0(x)\geq 0, \forall x\geq 0$, then the exponential term $\exp { \left(- W_{0} \left( O\left( n^{-1/{\alpha}}\right)\right) \right)}$ has values between 0 and 1. Note that as $n \rightarrow \infty$, the exponential term in the scaling law goes to 1. This implies that the exponential term is important to determine the scaling for $n$ small, while for large enough $n$ the upper bound is $O(n^{\frac{{\alpha}-1}{{\alpha}}})$. 
	
	If we consider $a(f) = 1$, i.e. the same path loss model as in \cite{GuptaKumar00}, and recall that $W_{0}(0) = 0$, then
\begin{eqnarray}
\lambda n \bar {L} \leq \frac{1}{\sqrt{\pi}}  {\left( \frac{2 \beta + 2}{ \beta} \right) }^{1/{\alpha}} W n^{\frac{{\alpha}-1}{{\alpha}}} 
\end{eqnarray}  
which is the original result of \cite{GuptaKumar00}. We have thus proved that the result in \cite{GuptaKumar00} is valid for $\alpha \geq 1$. 

\begin{figure}[t]
\centering	
\includegraphics[height=3.5in,width=3.5in,keepaspectratio]{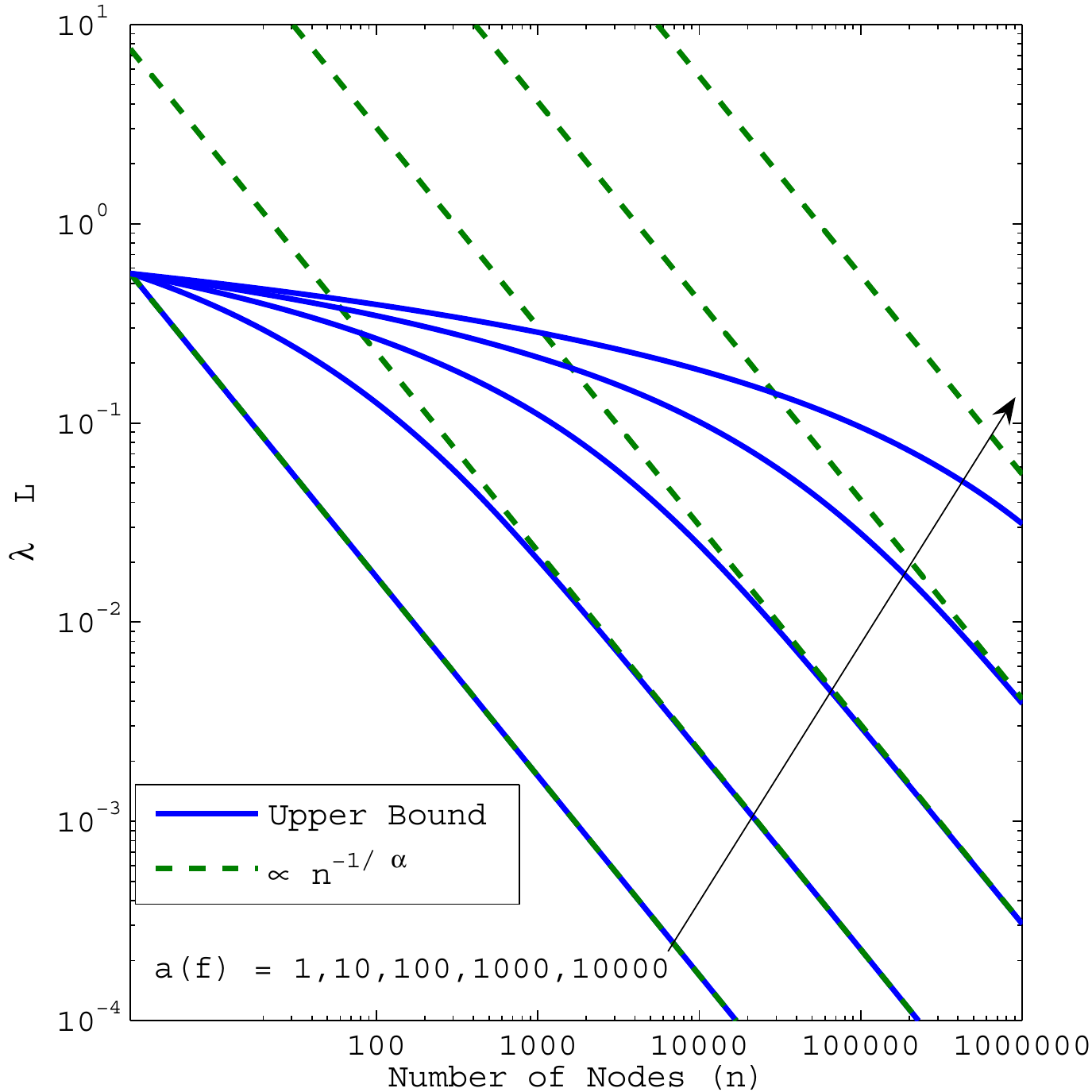}
\caption{Upper Bound on $\lambda \bar L$ for an arbitrarily chosen narrow band and different values of $a(f)$. Parameters used are $W=$~1~bps, $\alpha=$~1, $\beta = $~2, area = 1 km$^2$.}
\label{Upperbound.tag}
\end{figure}    

Fig. \ref{Upperbound.tag}
illustrates the upper bound on $\lambda \bar L$ for different values of $a(f)$ ranging from 1 to 10,000, which are characteristic of an underwater environment at different frequencies. For example, $a(f)=$~1,000 corresponds to a frequency of around 100~KHz. We have used $\alpha = $~1 and the parameters specified in the figure. We also plot dashed lines proportional to $n^{-1/\alpha}$.
As expected, as $n$ is large enough the exponential term of the upper bound becomes negligible making the bound scale as  $O(n^{-1/{\alpha}})$. However, for small values of $n$ the bound starts at a common point for the different $a(f)$ values, and decays very slowly. 
Fig. \ref{Upperbound.tag} also illustrates that the value of $a(f)$ determines the transition between these two operating regions: the larger $a(f)$, the greater $n$ has to be before transitioning. Of course, if we use a transmission band with high $a(f)$ each node will have to be able to transmit at higher power to reach its destination. In the underwater channel, it also means that a higher center frequency is required because $a(f)$ is an increasing function.

Finally, Fig. \ref{Upperbound.tag} shows that $\lambda \bar L$ remains almost constant for $n\leq 100$ nodes, $a(f) > 100$ and a disk area of 1 km$^2$, which corresponds to densities of up to 100 nodes per km$^2$. 
Note that the expected density of nodes in an underwater network is usually much lower given the applications for which they are deployed, e.g. environmental measurements. Thus, $\lambda \bar L$ is almost constant for practical purposes.      

\section{Low power - Narrow Band Case}

	As mentioned in the introduction, one of the characteristics of the underwater acoustic channel is that the optimal transmission band using the waterfilling principle depends strongly on the distance of a link. In particular, if the transmission power of a node is very low, then nodes will optimally transmit in different bands corresponding to different transmission distances. Thus, interference will come only from nodes transmitting in the same band. We have derived an expression for the power under these assumptions in Section II. In order to assign disjoint transmission bands, we divide the total transmission band of the system into non-overlapping bands of bandwidth $\Delta f$. We use $f_c(l)$ as the mapping between the transmission distance and the corresponding transmission band for a low-power/narrow-band scenario. Thus, if a node transmits to another node at a distance $l$, we assign the transmission band that contains the frequency $f_c(l)$ as in Fig.~\ref{Distance_Band.tag}. 

	The capacity analysis is similar as before if $a(f)$ is replaced by $a(f_{m})$ for each of the bands, where $f_m$ is the central frequency of transmission band $m$. 
	Let us assume that each node is capable of transmitting at $\triangle W$~bps in each band, where $\triangle W = \triangle f \log_{2} (1 + \beta)$, and $\triangle f$ is the bandwidth of each non-overlapping band. 
	
\subsection{Multi-Node Hopping}	
	Note that the definition of $H$ changes slightly when we allow multi-node hopping. In this case, $H \leq \frac{T|\Gamma| \triangle Wn}{2} = \frac{TWn}{2}$, where $\Gamma$ is the set of sub-bands used by the network, and $W = |\Gamma|\triangle W$.
	
For each of the different bands, the analysis is as before up to equation~\eqref{BreakingPoint}. At this point, let us define $\gamma_{\alpha} (f_{m})$ as $\gamma_{\alpha}$ for band $m$. Let us sum over all sub-bands and time slots to obtain
	
\begin{equation}
\sum_{s \in S} \sum_{m \in \Gamma } \sum _{i \in \tau }  {|X_{i} - X_{j(i)}|}^{\alpha} a(f_{m})^{|X_{i} - X_{j(i)}|} \notag
\end{equation} \begin{eqnarray}  \leq \frac{\beta + 1}{\beta} \triangle WT \sum_{m \in \Gamma} \gamma_{{\alpha}}(f_{m}) 
\end{eqnarray} 
where $S$ is the set of time slots. 
We can use the fact that $a(f_{m}) \geq a_{min}$, where $a_{min} = \min _{m \in \Gamma} a(f_{m})$. In the uncerwater scenario, $a_{min} = a(f_{min})$ because $a(f)$ is an increasing function. Defining $r^A (h,b,f_{min}) = {l(h,b)}^{\alpha} {a(f_{min})}^{l(h,b)}$, and following similar steps that lead to Eq.~\eqref{EQ11} we get

\begin{eqnarray}
\frac{1}{H}\sum_{b = 1} ^ {\lambda n T} \sum_{h = 1} ^ {h(b)} r^A (h,b,f_{min}) \leq \frac{\beta + 1}{\beta} \frac{\triangle WT}{H} \sum_{m \in \Gamma} \gamma_{{\alpha}}(f_{m}) \; . \notag
\end{eqnarray} 
Defining $\psi = (\ln{a(f_{min})})^{\alpha}  \frac{\beta + 1}{\beta} \frac{\triangle WT}{H} \sum_{m \in \Gamma} \gamma_{{\alpha}}(f_{m}) $, we can use a similar procedure as in the previous section, to show that the scaling law now becomes



\[
\lambda n\bar L \le \Phi Wn^{\frac{{\alpha - 1}}{\alpha}} \exp \left( { - W_0 \left( {\Phi \frac{{2\ln a(f_{\min } )}}{\alpha}\left( {\frac{1}{n}} \right)^{1/\alpha} } \right)} \right)
\]
where
\[
\Phi  = \frac{{2^{1/\alpha} }}{{\sqrt \pi  }}\left( {\frac{{\beta  + 1}}{\beta }} \right)^{1/\alpha} \left( {\frac{1}{{\left| \Gamma  \right|}}\sum\limits_{m \in \Gamma } {a(f{}_m)^{\frac{2}{{\sqrt \pi  }}} } } \right)^{1/\alpha} \; .
\]

The scaling law is similar in structure to the one obtained in Section III. However, the constant $\Phi$ depends on the average of a function of the absorption coefficients at $f_m, \forall m \in \Gamma$ instead of a particular value. Again, if $a(f) = 1, \forall f$ the result reduces to that of \cite{GuptaKumar00}.
 
\subsection{Direct Transmissions}
	If we constrain our system to perform direct transmissions only, using the fact that there is an assignment of frequency bands in terms of the distance, we can consider that $h(b) = 1, \forall b$, i.e. only one hop. Given the distance-band separation, the problem can be thought of as solving for several networks that lie on top of each other, in different layers with no cross-layer interference. Membership to the layers is based on the distance of the connection. In other words, each transmission band $m$ will have $n_{m}$ transmitters, where $n = \sum_{m \in \Gamma} n_{m}$ constitutes the total number of nodes in the network since each transmitter has only one intended destination.

	This causes a different capacity scaling for each of the transmission bands, i.e. the scaling for each transmission band will have the form of expression~\eqref{fixednarrowband} with $\triangle W$ instead of $W$ and $n_m$ instead of $n$ to obtain the scaling for band $m$. 

\section{High Power - Wide Band Case}
In this scenario, nodes have enough power to transmit in a wide transmission band $B$, which implies that the absorption cannot be considered to be a constant over the band. The band $B$ is again chosen using a waterfilling argument. We consider that the SINR requirement can depend on the frequency. That is
\begin{eqnarray}
\frac{ \frac{P_{i}(f)}{A(|X_{i} - X_{j(i)}|,f)}  }{ N(f) + \sum_{k \in \tau, k \neq i} \frac{P_{k}(f)}{A(|X_{k} - X_{j(i)}|,f) }} \geq \beta(f) \; .
\end{eqnarray} 

We define $W$ as the transmission data rate over the entire band, computed as
\begin{eqnarray} \label{W_lastcase}
W = \int_{f \in B} \log_{2} (1 + \beta(f)) df \; .
\end{eqnarray} 
If we assign a transmission rate to every sub-band $df$ of $dW = \log_{2} (1 + \beta(f)) df$, 
the analysis for each frequency is similar as in Section IV by letting $\triangle f \rightarrow 0$, renaming $\triangle f$ as $df$ and replacing the sums by integrals. Then, we have that
\begin{equation}
\frac{1}{H}\sum_{b = 1} ^ {\lambda n T} \sum_{h = 1} ^ {h(b)} W r^A (h,b,f_{min}) \leq  \frac{T}{H} \int_{W} \frac{\beta(f) + 1}{\beta(f)}\gamma_{{\alpha}}(f) dW\notag
\end{equation}\begin{equation}= \frac{T}{H} \int_{B} \frac{\beta(f) + 1}{\beta(f)}\gamma_{{\alpha}}(f) \log_{2} (1+\beta(f)) df\end{equation} 
where $f_{min} = \arg \min_f a(f)$ and $H$ can be shown to have the bound $H\leq \frac{TWn}{2}$ using the definition of $W$ in expression~\eqref{W_lastcase}.
Following the procedure of Section IV, we show that the scaling law for the high power - wide band case has the form 
\begin{eqnarray}
\lambda n\bar L \le \Theta Wn^{\frac{{\alpha - 1}}{\alpha}} \exp \left( { - W_0 \left( {\Theta \frac{{2\ln a(f_{\min } )}}{\alpha}\left( {\frac{1}{n}} \right)^{1/\alpha} } \right)} \right) \notag
\end{eqnarray}
where
\begin{eqnarray}
\Theta  = \frac{{2^{1/\alpha} }}{{\sqrt \pi  }}\left( {\frac{1}{W}\int\limits_B {\frac{{\left( {\beta (f) + 1} \right)a(f)^{\frac{2}{{\sqrt \pi  }}} \log _2 (1 + \beta (f))}}{{\beta (f)}}df} } \right)^{1/\alpha} \notag
\end{eqnarray}

\section{Conclusion}

This work presents upper bounds on the transport capacity of underwater acoustic networks with nodes deployed arbitrarily in a unit area disk. We study three cases of interest: an arbitrarily chosen narrow transmission band; the case of power limited nodes which transmit in disjoint narrow bands; and the case of nodes with high power capabilities that use of a wide transmission band. The choice of transmission band in the last two cases depends on the transmission distance and the physical characteristics of the channel, and is made in accordance with the waterfilling principle.

We have shown that the amount of information that can be exchanged by each source-destination pair in an underwater acoustic network goes to zero as the number of nodes $n$ goes to infinity, at least at rate $n^{-1/\alpha}e^{-W_0(O(n^{-1/\alpha}))}$. This rule is valid for the different scenarios in general, requiring only changes in the scaling constants. 
The throughput per source-destination pair has two different regions. For small $n$, the throughput decreases very slowly as $n$ increases. For large $n$, it decreases as $n^{-1/\alpha}$. Considering that $1 \leq\alpha\leq 2$ in an underwater channel, the available throughput for large $n$ decays more rapidly than in typical radio wireless networks. However, typical node densities in underwater correspond to the small $n$ regime. In a narrow band example with values of $a(f)$ characteristic of an underwater channel, we showed that the upper bound on the throughput remains almost constant for densities of less than 100 nodes per km$^2$. Most underwater networks have node densities in this range due to the applications for which they are deployed.   

Finally, we have pointed out some important characteristics of the underwater acoustic channel useful in future studies. For example, we could allow cooperation between nodes \textit{\`a la} Ozgur et al \cite{Ozgur07} taking advantage of the distance-band separation. That is, instead of performing time division between long and short transmissions, we could simply transmit in different bands that do not interfere with one another. This is important because acoustic transmissions have long propagation delays due to the speed of sound underwater ($\sim$1500 m/s), which reduces the usefulness of a time-division scheme.


\section*{Acknowledgment}

This work was supported in part by the National Science Foundation under grants No. 0520075, 0831728 and CNS-0627021, by ONR MURI Grant No. N00014-07-1-0738, and subcontract \# 060786 issued by BAE Systems National Security
Solutions, Inc. and supported by the Defense Advanced Research Projects
Agency (DARPA) and the Space and Naval Warfare System Center (SPAWARSYSCEN),
San Diego under Contract No. N66001-06-C-2020 (CBMANET).



%

\end{document}